\documentclass[10pt,conference]{IEEEtran}\IEEEoverridecommandlockouts
\usepackage{algorithmic}
\usepackage[linesnumbered,ruled]{algorithm2e}
\usepackage{amsfonts}
\usepackage{amsmath}
\usepackage{amssymb}
\usepackage{amsthm}
\usepackage{bm}
\usepackage{bbm}
\usepackage{booktabs}
\usepackage{caption}
\usepackage{CJK}
\usepackage{color}
\usepackage{cuted}
\usepackage{geometry}
\usepackage{graphicx}
\usepackage{indentfirst}
\usepackage[marginal]{footmisc}
\usepackage{multirow}
\usepackage{rotating}
\usepackage{subcaption}

\allowdisplaybreaks[4]

{\bgroup
 \addtolength\abovedisplayshortskip{#1}
 \addtolength\abovedisplayskip{#1}
 \addtolength\belowdisplayshortskip{#1}
 \addtolength\belowdisplayskip{#1}}
{\egroup\ignorespacesafterend}

\theoremstyle{plain}

\begin{document}

\title{A Flexible Design for Beam Squint Effect Suppression in IRS-Aided THz Communications}

\author{
\IEEEauthorblockN{Yanze Zhu$^{\mathrm{1}}$, Qingqing Wu$^{\mathrm{1}*}$, Wen Chen$^{\mathrm{1}}$, Yang Liu$^{\mathrm{2,3}}$, and Ruiqi Liu$^{\mathrm{4}}$}
\IEEEauthorblockA{1: Department of Electronic Engineering, Shanghai Jiao Tong University, Shanghai, China \\
2: School of Information and Communication Engineering, Dalian University of Technology, Dalian, China \\
3: National Mobile Communications Research Laboratory, Southeast University, Nanjing, China \\
4: Wireless and Computing Research Institute, ZTE Corporation, Shenzhen, China}
Emails: \{\{yanzezhu, qingqingwu, wenchen\}@sjtu.edu.cn, yangliu\_613@dlut.edu.cn, richie.leo@zte.com.cn\}
}

\maketitle

\begin{abstract}
In this paper, we study employing movable components on both base station (BS) and intelligent reflecting surface (IRS) in a wideband terahertz (THz) multiple-input-single-output (MISO) system, where the BS is equipped with a movable antenna (MA) array and the IRS consists of movable subarrays. To alleviate double beam squint effect caused by the coupling of beam squint at the BS and IRS, we propose to maximize the minimal received power across a wide THz spectrum by delicately configuring the positions of MAs and IRS subarrays, which is highly challenging. By adopting majorization-minimization (MM) methodology, we develop an algorithm to tackle the aforementioned optimization. Numerical results demonstrate the effectiveness of our proposed algorithm and the benefit of utilizing movable components on the BS and IRS to mitigate double beam squint effect in wideband THz communications.
\end{abstract}


\section{Introduction}

Terahertz (THz) communication has emerged as a promising technology for next-generation wireless networks [1]. Thanks to its unique feature of large bandwidth, THz communication is suitable for applications such as high-speed data transfer, imaging, and security scanning. However, due to severe path loss and atmospheric absorption, the potential of THz communication is bottlenecked in practice. \footnote{\hspace{0.35cm}* Corresponding author.} \footnote{\hspace{0.35cm}The work of Qingqing Wu was supported by National Key R\&D Program of China (2023YFB2905000), NSFC 62371289, Shanghai Jiao Tong University 2030 Initiative, and Guangdong science and technology program under grant 2022A0505050011. The work of Yang Liu was supported in part by the open research fund of National Mobile Communications Research Laboratory, Southeast University (No. 2025D06).}

To overcome this drawback, intelligent reflecting surface (IRS), which has captured great attention from both academia and industry, has been viewed as a promising technology to enhance the performance of THz communication [2]. IRS is equipped with a large number of low-cost passive reflecting elements which can intelligently manipulate the wireless signal propagation environment. By adjusting the phase of reflected signals, IRS can create favorable communication channels, thereby improving coverage, capacity, and energy efficiency (EE). The potentials of IRS have been greatly explored in the existing literature [3]-[5]. For instance, the authors of [3] uncovered that the transmit power at the base station (BS) can be reduced with the assistance of IRS. The work [4] considered EE maximization problem and demonstrated the effectiveness of IRS deployment for greatly boosting the system's EE. In [5], an IRS with true-time-delay (TTD) devices was deployed to enhance the sum-rate performance of THz communications.

Another promising solution is using movable antennas (MAs) [6], which represent a significant advancement in the field of wireless communications. Unlike traditional fixed position antennas (FPAs), MAs can be repositioned to optimize signal coverage, enhance communication reliability, and improve data throughput. A growing body of researches have concentrated on exploring the potentials of MA [7]-[9]. For instance, the authors of [7] modelled the channel of a single-input-single-output (SISO) MA system and theoretically proved that the signal-to-noise-ratio (SNR) performance can be dramatically improved with the employment of MAs. The work [8] investigated an MA-assisted wideband SISO system and verified the benefit of MA implementation for achievable rate improvement based on theoretical analysis and antenna position optimization. In [9], an MA array was deployed in a wideband millimeter-wave (mmWave) communication system to eliminate the beam squint effect.

Despite the aforementioned exciting progresses made in the existing literature, deploying movable components in IRS-aided wideband THz systems has not been thoroughly investigated. Note one prominent challenge of IRS-assisted wideband system is the \emph{double beam squint effect} [10], which is derived from the multiplicative nature of analog beamforming gain at the BS and IRS. Specifically, when subcarrier frequency significantly deviates from the reference tone (e.g., central frequency), the analog beamforming gain at the BS and IRS severely degrades, which further undermines the array gain of BS-IRS-user link. As unveiled by [9], beam squint effect can be remarkably alleviated by deploying MA array in wideband communication system. Inspired by the above observations, we are motivated to investigate movable component implementation at both the BS and IRS to effectively combat double beam squint effect in a wideband THz system, as shown in Fig. 1. To maintain low complexity and hardware cost at the IRS, the reflecting elements are grouped into multiple subarrays, each of which can flexibly move within the IRS surface. To the best of authors' knowledge, this is the first work studies double beam squint effect suppression strategy for wideband THz communication systems by deploying an MA array at the BS and utilizing movable subarrays on the IRS surface.

\section{System Model and Problem Formulation}

\subsection{System Model}

\begin{figure}[!t]
\centering
\includegraphics[scale=0.19]{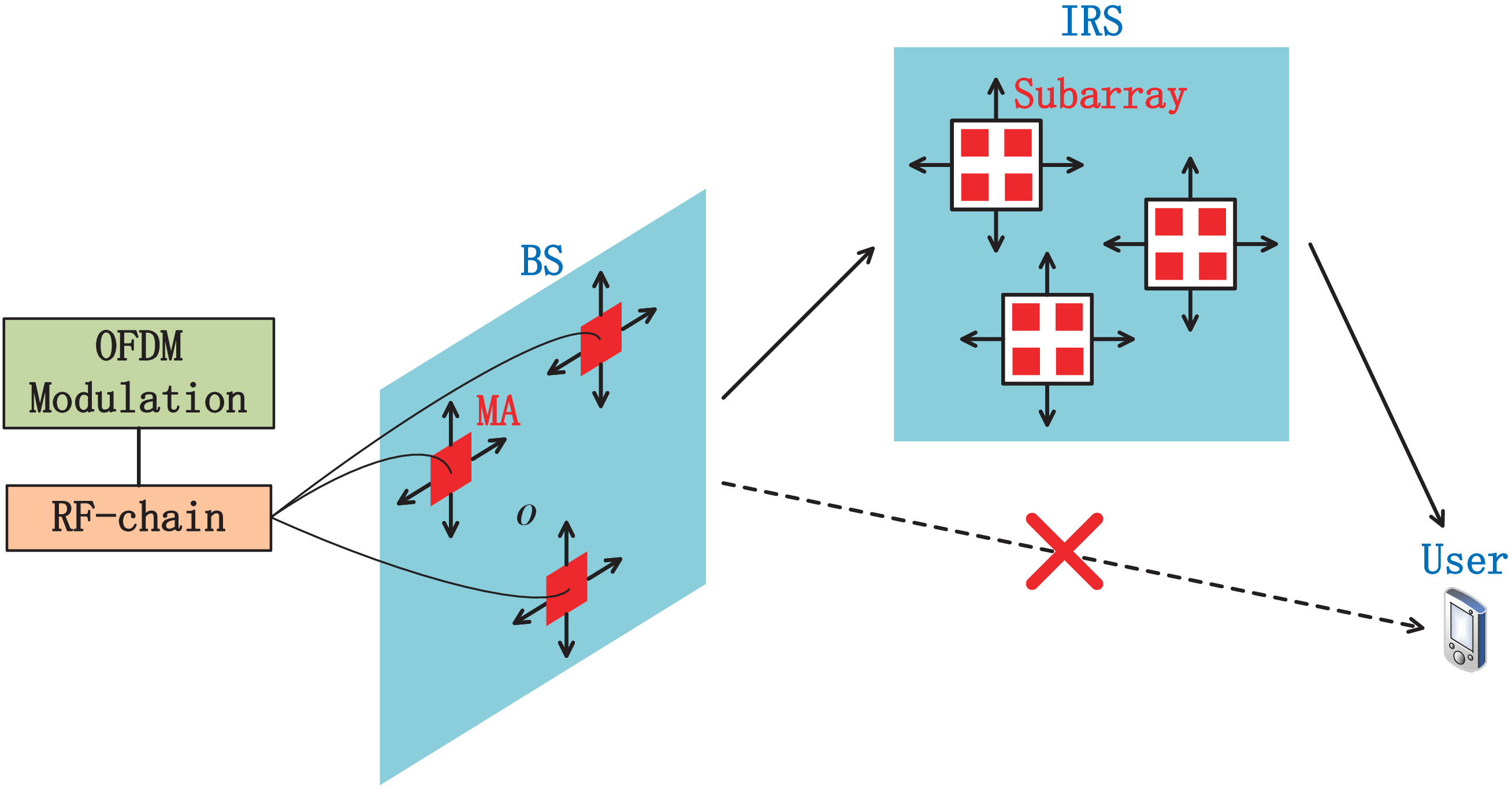}
\caption*{Fig. 1. Wideband MISO MA-enabled communication system aided by an IRS with movable subarrays.}
\end{figure}

As shown in Fig. 1, we consider an IRS-aided wideband THz multiple-input-single-output (MISO) communication system where a multi-antenna BS serves a single-antenna user with the assistance of an IRS. The BS is equipped with a 2-D planar array of size $A_{\mathrm{BS},1} \times A_{\mathrm{BS},2}$, which includes $M$ MAs, and the IRS is shaped as a 2-D planar surface of size $A_{\mathrm{IRS},1} \times A_{\mathrm{IRS},2}$. Denote $\mathcal{M} \triangleq \{ 1, \ldots, M \}$ as the set of MAs and $\mathbf{p}_{m}^{\mathrm{B}}, \; m \in \mathcal{M}$, as the relative position of the $m$-th MA with respect to (w.r.t.) the center of the BS array. Since all MAs should not move out of the BS array, each should guarantee $\mathbf{p}_{m}^{\mathrm{B}} \in \mathcal{C}_{\mathrm{BS}} \triangleq \{ (x, y)|x \in [-\frac{A_{\mathrm{BS},1}}{2}, \frac{A_{\mathrm{BS},1}}{2}], \; y \in [-\frac{A_{\mathrm{BS},2}}{2}, \frac{A_{\mathrm{BS},2}}{2}] \}, \; m \in \mathcal{M}$, where $\mathcal{C}_{\mathrm{BS}}$ represents the feasible moving region of the MAs.

The IRS elements are partitioned into $K$ subarrays, each of which can move flexibly on the IRS surface. Each subarray has the shape of uniform planar array (UPA) consisting of $J = J_{1}J_{2}$ reflecting elements, where $J_{1}$ and $J_{2}$ stand for the number of reflecting elements along its horizontal and vertical directions, respectively. Hence, the IRS has $N = KJ$ reflecting elements in total. Denote $\mathcal{K} \triangleq \{ 1, \ldots, K \}$ and $\mathcal{J} \triangleq \{ 1, \ldots, J \}$ as the sets of IRS subarrays and reflecting elements within one subarray, respectively. Furthermore, we define $\mathbf{c}_{k}^{\mathrm{R}}, \; k \in \mathcal{K}$, as the relative position of the $k$-th subarray's center w.r.t. the center of the IRS surface and $\mathbf{t}_{k,j}^{\mathrm{R}}, \; k \in \mathcal{K}, \; j \in \mathcal{J}$, as the relative position of the $j$-th reflecting element w.r.t. the center of the $k$-th subarray. Since all reflecting elements should maintain within the IRS surface, the constraint $\mathbf{c}_{k}^{\mathrm{R}} + \mathbf{t}_{k,j}^{\mathrm{R}} \in \mathcal{C}_{\mathrm{IRS}} \triangleq \{ (x, y)|x \in [-\frac{A_{\mathrm{IRS},1}}{2}, \frac{A_{\mathrm{IRS},1}}{2}], \; y \in [-\frac{A_{\mathrm{IRS},2}}{2}, \frac{A_{\mathrm{IRS},2}}{2}] \}, \; k \in \mathcal{K}, \; j \in \mathcal{J}$ should be satisfied, where $\mathcal{C}_{\mathrm{IRS}}$ represents the feasible dwelling region of the reflecting elements.

In this paper, we consider quasi-static far-field channel model. Since the attenuation of THz channels is severe, we assume that all channels only include line-of-sight (LoS) components. Besides, without loss of generality, the direct link between the BS and user is severely blocked. Denote $f_{l}, l \in \mathcal{L} \triangleq \{ 0, \ldots, L \}$, as the frequency of the $l$-th subcarrier with $f_{0} < \ldots < f_{L}$. Then, the channel between the BS and IRS at the $l$-th subcarrier is given by
\begin{align}
\mathbf{G}_{l}(\tilde{\mathbf{p}}_{\mathrm{B}}, \tilde{\mathbf{c}}_{\mathrm{R}}) &= \alpha_{\mathrm{G},l}e^{-j2\pi\tau_{\mathrm{G}}f_{l}}\mathbf{a}(\tilde{\mathbf{c}}_{\mathrm{R}}, f_{l}, \vartheta_{\mathrm{R}}^{\mathrm{r}}, \varphi_{\mathrm{R}}^{\mathrm{r}}) \notag\\
& \times \mathbf{b}^{\mathsf{H}}(\tilde{\mathbf{p}}_{\mathrm{B}}, f_{l}, \vartheta_{\mathrm{B}}, \varphi_{\mathrm{B}}), \; l \in \mathcal{L},
\end{align}
where $\tilde{\mathbf{p}}_{\mathrm{B}} \triangleq [(\mathbf{p}_{1}^{\mathrm{B}})^{\mathsf{T}}, \ldots, (\mathbf{p}_{M}^{\mathrm{B}})^{\mathsf{T}}]^{\mathsf{T}}$ and $\tilde{\mathbf{c}}_{\mathrm{R}} \triangleq [(\mathbf{c}_{1}^{\mathrm{R}})^{\mathsf{T}}, \ldots, (\mathbf{c}_{K}^{\mathrm{R}})^{\mathsf{T}}]^{\mathsf{T}}$ represent the position vectors of MAs and IRS subarrays, respectively. $\tau_{\mathrm{G}}$ stands for the path delay of BS-IRS channel. $\alpha_{\mathrm{G},l}$ represents the path loss of BS-IRS channel at the $l$-th subcarrier, which can be expressed as
\begin{align}
\alpha_{\mathrm{G},l} = \frac{c}{4\pi f_{l}d_{\mathrm{G}}}e^{-\frac{1}{2}\kappa_{\mathrm{abs}}(f_{l})d_{\mathrm{G}}},
\end{align}
where $c$, $d_{\mathrm{G}}$, and $\kappa_{\mathrm{abs}}(f_{l})$ are the speed of light, the distance between the BS and IRS, and the molecular absorption factor, respectively. $\mathbf{a}(\tilde{\mathbf{c}}_{\mathrm{R}}, f_{l}, \vartheta_{\mathrm{R}}^{\mathrm{r}}, \varphi_{\mathrm{R}}^{\mathrm{r}})$ indicates the steering vector at the IRS, which is given by
\begin{align}
\mathbf{a}(\tilde{\mathbf{c}}_{\mathrm{R}}, f_{l}, \vartheta_{\mathrm{R}}^{\mathrm{r}}, \varphi_{\mathrm{R}}^{\mathrm{r}}) = [&\mathbf{a}_{1}^{\mathsf{T}}(\mathbf{c}_{1}^{\mathrm{R}}, f_{l}, \vartheta_{\mathrm{R}}^{\mathrm{r}}, \varphi_{\mathrm{R}}^{\mathrm{r}}), \notag\\
&\ldots, \mathbf{a}_{K}^{\mathsf{T}}(\mathbf{c}_{K}^{\mathrm{R}}, f_{l}, \vartheta_{\mathrm{R}}^{\mathrm{r}}, \varphi_{\mathrm{R}}^{\mathrm{r}})]^{\mathsf{T}},
\end{align}
where $\mathbf{a}_{k}(\mathbf{c}_{k}^{\mathrm{R}}, f_{l}, \vartheta_{\mathrm{R}}^{\mathrm{r}}, \varphi_{\mathrm{R}}^{\mathrm{r}}), \; k \in \mathcal{K}$, represents the steering vector at the $k$-th IRS subarray, which reads
\begin{align}
\mathbf{a}_{k}(\mathbf{c}_{k}^{\mathrm{R}}, f_{l}, \vartheta_{\mathrm{R}}^{\mathrm{r}}, \varphi_{\mathrm{R}}^{\mathrm{r}}) = [&e^{j\frac{2\pi f_{l}}{c}(\mathbf{c}_{k}^{\mathrm{R}} + \mathbf{t}_{k,1}^{\mathrm{R}})^{\mathsf{T}}\bm{\rho}_{\mathrm{R}}^{\mathrm{r}}}, \notag\\
&\ldots, e^{j\frac{2\pi f_{l}}{c}(\mathbf{c}_{k}^{\mathrm{R}} + \mathbf{t}_{k,J}^{\mathrm{R}})^{\mathsf{T}}\bm{\rho}_{\mathrm{R}}^{\mathrm{r}}}]^{\mathsf{T}},
\end{align}
where $\bm{\rho}_{\mathrm{R}}^{\mathrm{r}} \triangleq [\mathsf{sin}\vartheta_{\mathrm{R}}^{\mathrm{r}}\mathsf{sin}\varphi_{\mathrm{R}}^{\mathrm{r}}, \mathsf{cos}\varphi_{\mathrm{R}}^{\mathrm{r}}]^{\mathsf{T}}$, with $\vartheta_{\mathrm{R}}^{\mathrm{r}}$ and $\varphi_{\mathrm{R}}^{\mathrm{r}}$ being the azimuth and elevation angles of angle-of-arrival (AoA) at the IRS, respectively$^{1}$\footnote{$^{1}$Note in this paper, all the mentioned azimuth and elevation angles are assumed to be known, which can be estimated by various existing sensing techniques, e.g., [11], [12].}. $\mathbf{b}(\tilde{\mathbf{p}}_{\mathrm{B}}, f_{l}, \vartheta_{\mathrm{B}}, \varphi_{\mathrm{B}})$ stands for the steering vector at the BS, which can be written as
\begin{align}
\mathbf{b}(\tilde{\mathbf{p}}_{\mathrm{B}}, f_{l}, \vartheta_{\mathrm{B}}, \varphi_{\mathrm{B}}) \!\!=\!\! [e^{j\frac{2\pi f_{l}}{c}(\mathbf{p}_{1}^{\mathrm{B}})^{\mathsf{T}}\bm{\rho}_{\mathrm{B}}}, \ldots, e^{j\frac{2\pi f_{l}}{c}(\mathbf{p}_{M}^{\mathrm{B}})^{\mathsf{T}}\bm{\rho}_{\mathrm{B}}}]^{\mathsf{T}},
\end{align}
where $\bm{\rho}_{\mathrm{B}} \triangleq [\mathsf{cos}\vartheta_{\mathrm{B}}\mathsf{sin}\varphi_{\mathrm{B}}, \mathsf{cos}\varphi_{\mathrm{B}}]^{\mathsf{T}}$, with $\vartheta_{\mathrm{B}}$ and $\varphi_{\mathrm{B}}$ being the azimuth and elevation angles of angle-of-departure (AoD) at the BS, respectively.

Similarly, the channel between the IRS and user at the $l$-th subcarrier is given as follows
\begin{align}
\mathbf{h}_{l}(\tilde{\mathbf{c}}_{\mathrm{R}}) = \alpha_{\mathrm{h},l}e^{-j2\pi\tau_{\mathrm{h}}f_{l}}\mathbf{a}(\tilde{\mathbf{c}}_{\mathrm{R}}, f_{l}, \vartheta_{\mathrm{R}}^{\mathrm{t}}, \varphi_{\mathrm{R}}^{\mathrm{t}}), \; l \in \mathcal{L},
\end{align}
where $\tau_{\mathrm{h}}$ represents the path delay of IRS-user channel. $\vartheta_{\mathrm{R}}^{\mathrm{t}}$ and $\varphi_{\mathrm{R}}^{\mathrm{t}}$ are the azimuth and elevation angles of AoD at the IRS, respectively. $\alpha_{\mathrm{h},l}$ indicates the path loss of IRS-user channel at the $l$-th subcarrier, which can be expressed as
\begin{align}
\alpha_{\mathrm{h},l} = \frac{c}{4\pi f_{l}d_{\mathrm{h}}}e^{-\frac{1}{2}\kappa_{\mathrm{abs}}(f_{l})d_{\mathrm{h}}},
\end{align}
where $d_{\mathrm{h}}$ stands for the distance between the IRS and user.

The received signal of the user at the $l$-th subcarrier is given by
\begin{align}
y_{l} = \mathbf{h}_{l}^{\mathsf{H}}(\tilde{\mathbf{c}}_{\mathrm{R}})\bm{\Theta}\mathbf{G}_{l}(\tilde{\mathbf{p}}_{\mathrm{B}}, \tilde{\mathbf{c}}_{\mathrm{R}})\mathbf{f}s_{l} + n_{l}, \; l \in \mathcal{L},
\end{align}
where $s_{l}$ and $n_{l}$ are the transmit signal at the BS and the thermal noise at the user, respectively, with $\mathbb{E} \{ s_{l} \} = 0$ and $\mathbb{E} \{ |s_{l}|^{2} \} = 1$, $\mathbf{f}$ and $\bm{\Theta}$ represent the BS beamforming vector and IRS phase shift matrix, respectively. For simplicity, we assume that $s_{l}$ and $n_{l}$, for all $l \in \mathcal{L}$, are uncorrelated [3].

According to (8), the amplitude of user's received signal at the $l$-th subcarrier reads$^{2}$\footnote{$^{2}$Note in the following, the amplitude and power of received signals do not include thermal noise.}
\begin{align}
\mathsf{g}_{l}(\tilde{\mathbf{p}}_{\mathrm{B}}, \tilde{\mathbf{c}}_{\mathrm{R}}) = |\mathbf{h}_{l}^{\mathsf{H}}(\tilde{\mathbf{c}}_{\mathrm{R}})\bm{\Theta}\mathbf{G}_{l}(\tilde{\mathbf{p}}_{\mathrm{B}}, \tilde{\mathbf{c}}_{\mathrm{R}})\mathbf{f}|, \; l \in \mathcal{L}.
\end{align}

\subsection{Double Beam Squint Effect}

To investigate the impact of wide spectrum on the amplitude of user's received signal shown by (9), we adopt the beamformers based on central frequency $f_{\mathrm{c}}$ at the BS and IRS [13], which are respectively given by
\begin{align}
\mathbf{f}(\tilde{\mathbf{p}}_{\mathrm{B}}) &\!=\! \mathbf{b}(\tilde{\mathbf{p}}_{\mathrm{B}}, f_{\mathrm{c}}, \vartheta_{\mathrm{B}}, \varphi_{\mathrm{B}}), \\
\bm{\Theta}(\tilde{\mathbf{c}}_{\mathrm{R}}) &\!=\! \mathsf{Diag}(\mathbf{a}(\tilde{\mathbf{c}}_{\mathrm{R}}, f_{\mathrm{c}}, \vartheta_{\mathrm{R}}^{\mathrm{t}}, \varphi_{\mathrm{R}}^{\mathrm{t}}) \!\odot\! \mathbf{a}^{*}(\tilde{\mathbf{c}}_{\mathrm{R}}, f_{\mathrm{c}}, \vartheta_{\mathrm{R}}^{\mathrm{r}}, \varphi_{\mathrm{R}}^{\mathrm{r}})).
\end{align}
Besides, based on (9), $\mathsf{g}_{l}(\tilde{\mathbf{p}}_{\mathrm{B}}, \tilde{\mathbf{c}}_{\mathrm{R}}), \; l \in \mathcal{L}$, can be further casted as
\begin{align}
\mathsf{g}_{l}(\tilde{\mathbf{p}}_{\mathrm{B}}, \tilde{\mathbf{c}}_{\mathrm{R}}) &= \alpha_{\mathrm{G},l}\alpha_{\mathrm{h},l}|\mathbf{a}^{\mathsf{H}}(\tilde{\mathbf{c}}_{\mathrm{R}}, f_{l}, \vartheta_{\mathrm{R}}^{\mathrm{t}}, \varphi_{\mathrm{R}}^{\mathrm{t}})\bm{\Theta}(\tilde{\mathbf{c}}_{\mathrm{R}}) \notag\\
& \times \mathbf{a}(\tilde{\mathbf{c}}_{\mathrm{R}}, f_{l}, \vartheta_{\mathrm{R}}^{\mathrm{r}}, \varphi_{\mathrm{R}}^{\mathrm{r}})||\mathbf{b}^{\mathsf{H}}(\tilde{\mathbf{p}}_{\mathrm{B}}, f_{l}, \vartheta_{\mathrm{B}}, \varphi_{\mathrm{B}})\mathbf{f}(\tilde{\mathbf{p}}_{\mathrm{B}})| \notag\\
& = \alpha_{\mathrm{G},l}\alpha_{\mathrm{h},l}\mathsf{g}_{l}^{\mathrm{R}}(\tilde{\mathbf{c}}_{\mathrm{R}})\mathsf{g}_{l}^{\mathrm{B}}(\tilde{\mathbf{p}}_{\mathrm{B}}),
\end{align}
where $\mathsf{g}_{l}^{\mathrm{R}}(\tilde{\mathbf{c}}_{\mathrm{R}}) \triangleq |\mathbf{a}^{\mathsf{H}}(\tilde{\mathbf{c}}_{\mathrm{R}}, f_{l}, \vartheta_{\mathrm{R}}^{\mathrm{t}}, \varphi_{\mathrm{R}}^{\mathrm{t}})\bm{\Theta}(\tilde{\mathbf{c}}_{\mathrm{R}})\mathbf{a}(\tilde{\mathbf{c}}_{\mathrm{R}}, f_{l}, \vartheta_{\mathrm{R}}^{\mathrm{r}}, \varphi_{\mathrm{R}}^{\mathrm{r}})|$ and $\mathsf{g}_{l}^{\mathrm{B}}(\tilde{\mathbf{p}}_{\mathrm{B}}) \triangleq |\mathbf{b}^{\mathsf{H}}(\tilde{\mathbf{p}}_{\mathrm{B}}, f_{l}, \vartheta_{\mathrm{B}}, \varphi_{\mathrm{B}})\mathbf{f}(\tilde{\mathbf{p}}_{\mathrm{B}})|$.

As indicated by (12), the amplitude of user's received signal is related to the subcarrier frequency $f_{l}$. For wideband system, $f_{l}$ significantly deviates from $f_{\mathrm{c}}$ for lots of $l$'s, which lead to the array gain of the BS and that of the IRS severely fluctuating across subcarriers. This phenomenon is the so-called beam squint effect at the BS and that at the IRS. Besides, due to the multiplicative nature of $\mathsf{g}_{l}^{\mathrm{R}}(\tilde{\mathbf{c}}_{\mathrm{R}})$ and $\mathsf{g}_{l}^{\mathrm{B}}(\tilde{\mathbf{p}}_{\mathrm{B}})$ according to (12), the amplitude of user's received signal $\mathsf{g}_{l}(\tilde{\mathbf{p}}_{\mathrm{B}}, \tilde{\mathbf{c}}_{\mathrm{R}})$ further suffers from more severe degradation across subcarriers, which is the so-called double beam squint effect [10], [13].

\subsection{Problem Formulation}

To alleviate the aforementioned double beam squint effect and balance the communication rate of data stream among different subcarriers, in this paper, we propose to maximize the minimal received power among the entire frequency band by jointly configuring MAs within the BS and subarrays within the IRS, whose optimization problem is formulated as
\begin{align}
(\mathcal{P}1): \max_{\tilde{\mathbf{p}}_{\mathrm{B}}, \tilde{\mathbf{c}}_{\mathrm{R}}} &\min_{l \in \mathcal{L}} \; \mathsf{h}_{l}(\tilde{\mathbf{p}}_{\mathrm{B}}, \tilde{\mathbf{c}}_{\mathrm{R}}) \\
\mathrm{s.t.} \; &\mathbf{p}_{m}^{\mathrm{B}} \in \mathcal{C}_{\mathrm{BS}}, \; m \in \mathcal{M}, \tag{13a}\\
&\| \mathbf{p}_{m}^{\mathrm{B}} - \mathbf{p}_{s}^{\mathrm{B}} \|_{2} \geq D_{\mathrm{BS}}, \; m, s \in \mathcal{M}, \; m \neq s, \tag{13b}\\
&\mathbf{c}_{k}^{\mathrm{R}} + \mathbf{t}_{k,j}^{\mathrm{R}} \in \mathcal{C}_{\mathrm{IRS}}, \; k \in \mathcal{K}, \; j \in \mathcal{J}, \tag{13c}\\
&\| \mathbf{c}_{k}^{\mathrm{R}} - \mathbf{c}_{i}^{\mathrm{R}} \|_{2} \geq D_{\mathrm{IRS}}, \; k, i \in \mathcal{K}, \; k \neq i, \tag{13d}
\end{align}
where $\mathsf{h}_{l}(\tilde{\mathbf{p}}_{\mathrm{B}}, \tilde{\mathbf{c}}_{\mathrm{R}}) \triangleq \mathsf{g}_{l}^{2}(\tilde{\mathbf{p}}_{\mathrm{B}}, \tilde{\mathbf{c}}_{\mathrm{R}})$, constraints (13a) and (13c) indicate that each MA at the BS and each subarray at the IRS should move within the feasible region $\mathcal{C}_{\mathrm{BS}}$ and $\mathcal{C}_{\mathrm{IRS}}$, i.e., the BS antenna array and the IRS surface, respectively. Constraints (13b) and (13d) describe that each MA at the BS and each subarray at the IRS should maintain a distance of at least $D_{\mathrm{BS}}$ and $D_{\mathrm{IRS}}$ from other MAs/subarrays while moving to avoid antenna coupling and subarray collision, respectively. Problem $(\mathcal{P}1)$ is highly nonconvex due to its objective and constraints (13b) and (13d), which is challenging to solve. In the following, we develop an algorithm to attack $(\mathcal{P}1)$.

\section{Proposed Algorithm}

In this section, we propose an algorithm to tackle $(\mathcal{P}1)$. First, via introducing a slack variable, $(\mathcal{P}1)$ can be equivalently transformed into the following more tractable form
\begin{align}
(\mathcal{P}2): &\max_{\tilde{\mathbf{p}}_{\mathrm{B}}, \tilde{\mathbf{c}}_{\mathrm{R}}, \kappa} \; \kappa \\
\mathrm{s.t.} \; &\mathsf{h}_{l}(\tilde{\mathbf{p}}_{\mathrm{B}}, \tilde{\mathbf{c}}_{\mathrm{R}}) \geq \kappa, \; l \in \mathcal{L}, \tag{14a}\\
&\mathbf{p}_{m}^{\mathrm{B}} \in \mathcal{C}_{\mathrm{BS}}, \; m \in \mathcal{M}, \tag{14b}\\
&\| \mathbf{p}_{m}^{\mathrm{B}} - \mathbf{p}_{s}^{\mathrm{B}} \|_{2} \geq D_{\mathrm{BS}}, \; m, s \in \mathcal{M}, \; m \neq s, \tag{14c}\\
&\mathbf{c}_{k}^{\mathrm{R}} + \mathbf{t}_{k,j}^{\mathrm{R}} \in \mathcal{C}_{\mathrm{IRS}}, \; k \in \mathcal{K}, \; j \in \mathcal{J}, \tag{14d}\\
&\| \mathbf{c}_{k}^{\mathrm{R}} - \mathbf{c}_{i}^{\mathrm{R}} \|_{2} \geq D_{\mathrm{IRS}}, \; k, i \in \mathcal{K}, \; k \neq i. \tag{14e}
\end{align}

In the following, we utilize block coordinate descent (BCD) method to handle $(\mathcal{P}2)$ by alternatively optimizing one MA's or subarray's position at a time while keeping others being fixed. Specifically, taking the antenna position of each MA at the BS into account first, the optimization w.r.t. the $m$-th MA's position $\mathbf{p}_{m}^{\mathrm{B}}$ is given by
\begin{align}
(\mathcal{P}3_{m}): \max_{\mathbf{p}_{m}^{\mathrm{B}}, \kappa} \; &\kappa \\
\mathrm{s.t.} \; &\mathsf{h}_{l}(\mathbf{p}_{m}^{\mathrm{B}}) \geq \kappa, \; l \in \mathcal{L}, \tag{15a}\\
&\mathbf{p}_{m}^{\mathrm{B}} \in \mathcal{C}_{\mathrm{BS}}, \tag{15b}\\
&\| \mathbf{p}_{m}^{\mathrm{B}} - \mathbf{p}_{s}^{\mathrm{B}} \|_{2} \geq D_{\mathrm{BS}}, \; s \in \mathcal{M}, \; m \neq s, \tag{15c}
\end{align}
which is still difficult to solve since constraints (15a) and (15c) are nonconvex. To overcome this difficulty, we exploit majorization-minimization (MM) [14] method to convexify (15a) and (15c), respectively.

Specifically, considering (15a) first, due to the fact that $\mathsf{h}_{l}(\mathbf{p}_{m}^{\mathrm{B}})$ is neither convex nor concave w.r.t. $\mathbf{p}_{m}^{\mathrm{B}}$, we examine its second-order Taylor expansion w.r.t. $\mathbf{p}_{m}^{\mathrm{B}}$ at the point $\hat{\mathbf{p}}_{m}^{\mathrm{B}}$, and replace the Hessian matrix by a smaller one to construct a tight globally lower-bound surrogate for $\mathsf{h}_{l}(\mathbf{p}_{m}^{\mathrm{B}})$. Due to space limited, we directly present the surrogate function of $\mathsf{h}_{l}(\mathbf{p}_{m}^{\mathrm{B}})$, which is given as follows
\begin{align}
\mathsf{h}_{l}(\mathbf{p}_{m}^{\mathrm{B}}|\hat{\mathbf{p}}_{m}^{\mathrm{B}}) &= \mathsf{h}_{l}(\hat{\mathbf{p}}_{m}^{\mathrm{B}}) + \nabla_{\mathbf{p}_{m}^{\mathrm{B}}}^{\mathsf{T}}\mathsf{h}_{l}(\mathbf{p}_{m}^{\mathrm{B}})|_{\mathbf{p}_{m}^{\mathrm{B}} = \hat{\mathbf{p}}_{m}^{\mathrm{B}}}(\mathbf{p}_{m}^{\mathrm{B}} - \hat{\mathbf{p}}_{m}^{\mathrm{B}}) \notag\\
& + (\mathbf{p}_{m}^{\mathrm{B}} - \hat{\mathbf{p}}_{m}^{\mathrm{B}})^{\mathsf{T}}\frac{\mathbf{M}_{m,l}^{\mathrm{B}}}{2}(\mathbf{p}_{m}^{\mathrm{B}} - \hat{\mathbf{p}}_{m}^{\mathrm{B}}),
\end{align}
where
\begin{align}
&\nabla_{\mathbf{p}_{m}^{\mathrm{B}}}\mathsf{h}_{l}(\mathbf{p}_{m}^{\mathrm{B}}) = \bigg[\frac{\partial\mathsf{h}_{l}(\mathbf{p}_{m}^{\mathrm{B}})}{\partial[\mathbf{p}_{m}^{\mathrm{B}}]_{1}}, \frac{\partial\mathsf{h}_{l}(\mathbf{p}_{m}^{\mathrm{B}})}{\partial[\mathbf{p}_{m}^{\mathrm{B}}]_{2}}\bigg]^{\mathsf{T}}, \\
&\frac{\partial\mathsf{h}_{l}(\mathbf{p}_{m}^{\mathrm{B}})}{\partial[\mathbf{p}_{m}^{\mathrm{B}}]_{1}} = -2\beta_{l}^{\mathrm{B}}|C_{m,l}^{\mathrm{B}}|F_{l}\mathsf{cos}\vartheta_{\mathrm{B}}\mathsf{sin}\varphi_{\mathrm{B}} \notag\\
&\qquad\qquad \times \mathsf{sin}(F_{l}(\mathbf{p}_{m}^{\mathrm{B}})^{\mathsf{T}}\bm{\rho}_{\mathrm{B}} - \angle(C_{m,l}^{\mathrm{B}})), \\
&\frac{\partial\mathsf{h}_{l}(\mathbf{p}_{m}^{\mathrm{B}})}{\partial[\mathbf{p}_{m}^{\mathrm{B}}]_{2}} = -2\beta_{l}^{\mathrm{B}}|C_{m,l}^{\mathrm{B}}|F_{l}\mathsf{cos}\varphi_{\mathrm{B}} \notag\\
&\qquad\qquad \times \mathsf{sin}(F_{l}(\mathbf{p}_{m}^{\mathrm{B}})^{\mathsf{T}}\bm{\rho}_{\mathrm{B}} - \angle(C_{m,l}^{\mathrm{B}})), \\
&\mathbf{M}_{m,l}^{\mathrm{B}} = -2\beta_{l}^{\mathrm{B}}|C_{m,l}^{\mathrm{B}}|F_{l}^{2}(\mathsf{Diag}([(\mathsf{cos}\vartheta_{\mathrm{B}}\mathsf{sin}\varphi_{\mathrm{B}})^{2}, (\mathsf{cos}\varphi_{\mathrm{B}})^{2}]^{\mathsf{T}}) \notag\\
&\qquad\;\; + |\mathsf{cos}\vartheta_{\mathrm{B}}\mathsf{sin}\varphi_{\mathrm{B}}\mathsf{cos}\varphi_{\mathrm{B}}|\mathbf{I}_{2}),
\end{align}
with $F_{l} \triangleq \frac{2\pi(f_{\mathrm{c}} - f_{l})}{c}$, $\beta_{l}^{\mathrm{B}} \triangleq \alpha_{\mathrm{G},l}^{2}\alpha_{\mathrm{h},l}^{2}(\mathsf{g}_{l}^{\mathrm{R}}(\tilde{\mathbf{c}}_{\mathrm{R}}))^{2}$, $C_{m,l}^{\mathrm{B}} \triangleq \sum_{n = 1, n \neq m}^{M}e^{jF_{l}(\mathbf{p}_{n}^{\mathrm{B}})^{\mathsf{T}}\bm{\rho}_{\mathrm{B}}}$, $\mathbf{I}_{2}$ being an identity matrix of dimension $2 \times 2$ and $\angle(\cdot)$ picking up the phase of input complex scalar. Please refer to our journal version [15] for more details.

In the following, we resolve the non-convexity of (15c), which is much easier because the left hand side of (15c) is convex such that its first-order Taylor expansion serves as a globally concave lower bound. Therefore, by leveraging the following first-order Taylor expansion
\begin{align}
\| \mathbf{p}_{m}^{\mathrm{B}} - \mathbf{p}_{s}^{\mathrm{B}} \|_{2} \geq \frac{(\hat{\mathbf{p}}_{m}^{\mathrm{B}} - \mathbf{p}_{s}^{\mathrm{B}})^{\mathsf{T}}}{\| \hat{\mathbf{p}}_{m}^{\mathrm{B}} - \mathbf{p}_{s}^{\mathrm{B}} \|_{2}}(\mathbf{p}_{m}^{\mathrm{B}} - \mathbf{p}_{s}^{\mathrm{B}}),
\end{align}
the convexification for (15c) can be achieved.

Based on the above two developed concave lower bounds (16) and (21), a convex optimization problem can be acquired, given as follows
\begin{align}
&(\mathcal{P}4_{m}): \max_{\mathbf{p}_{m}^{\mathrm{B}}, \kappa} \; \kappa \\
\mathrm{s.t.} \; &\mathsf{h}_{l}(\mathbf{p}_{m}^{\mathrm{B}}|\hat{\mathbf{p}}_{m}^{\mathrm{B}}) \geq \kappa, \; l \in \mathcal{L}, \tag{22a}\\
&\mathbf{p}_{m}^{\mathrm{B}} \in \mathcal{C}_{\mathrm{BS}}, \tag{22b}\\
&\frac{(\hat{\mathbf{p}}_{m}^{\mathrm{B}} - \mathbf{p}_{s}^{\mathrm{B}})^{\mathsf{T}}}{\| \hat{\mathbf{p}}_{m}^{\mathrm{B}} - \mathbf{p}_{s}^{\mathrm{B}} \|_{2}}(\mathbf{p}_{m}^{\mathrm{B}} - \mathbf{p}_{s}^{\mathrm{B}}) \geq D_{\mathrm{BS}}, \; s \!\in\! \mathcal{M}, \; m \!\neq\! s, \tag{22c}
\end{align}
which can be solved by existing standard numerical solvers, e.g., CVX.

Next, we proceed to optimize the position of each IRS subarray with other variables being fixed. The subproblem w.r.t. the $k$-th subarray's position $\mathbf{c}_{k}^{\mathrm{R}}$ can be expressed as
\begin{align}
(\mathcal{P}5_{k}): \max_{\mathbf{c}_{k}^{\mathrm{R}}, \kappa} \; &\kappa \\
\mathrm{s.t.} \; &\mathsf{h}_{l}(\mathbf{c}_{k}^{\mathrm{R}}) \geq \kappa, \; l \in \mathcal{L}, \tag{23a}\\
&\mathbf{c}_{k}^{\mathrm{R}} + \mathbf{t}_{k,j}^{\mathrm{R}} \in \mathcal{C}_{\mathrm{IRS}}, \; j \in \mathcal{J}, \tag{23b}\\
&\| \mathbf{c}_{k}^{\mathrm{R}} - \mathbf{c}_{i}^{\mathrm{R}} \|_{2} \geq D_{\mathrm{IRS}}, \; i \in \mathcal{K}, \; k \neq i, \tag{23c}
\end{align}
which is non-convex due to constraints (23a) and (23c). Taking (23a) into consideration first, similar to the procedure of convexifying (15a), we tackle (23a) via adopting the second-order Taylor expansion of $\mathsf{h}_{l}(\mathbf{c}_{k}^{\mathrm{R}})$ w.r.t. $\mathbf{c}_{k}^{\mathrm{R}}$ at the point $\hat{\mathbf{c}}_{k}^{\mathrm{R}}$ due to non-convexity and non-concavity of $\mathsf{h}_{l}(\mathbf{c}_{k}^{\mathrm{R}})$, and surrogating its Hessian matrix by a smaller one which is negative semidefinite. Here we still directly provide the surrogate function of $\mathsf{h}_{l}(\mathbf{c}_{k}^{\mathrm{R}})$, which is given by
\begin{align}
\mathsf{h}_{l}(\mathbf{c}_{k}^{\mathrm{R}}|\hat{\mathbf{c}}_{k}^{\mathrm{R}}) &= \mathsf{h}_{l}(\hat{\mathbf{c}}_{k}^{\mathrm{R}}) + \nabla_{\mathbf{c}_{k}^{\mathrm{R}}}^{\mathsf{T}}\mathsf{h}_{l}(\mathbf{c}_{k}^{\mathrm{R}})|_{\mathbf{c}_{k}^{\mathrm{R}} = \hat{\mathbf{c}}_{k}^{\mathrm{R}}}(\mathbf{c}_{k}^{\mathrm{R}} - \hat{\mathbf{c}}_{k}^{\mathrm{R}}) \notag\\
& + (\mathbf{c}_{k}^{\mathrm{R}} - \hat{\mathbf{c}}_{k}^{\mathrm{R}})^{\mathsf{T}}\frac{\mathbf{M}_{k,l}^{\mathrm{R}}}{2}(\mathbf{c}_{k}^{\mathrm{R}} - \hat{\mathbf{c}}_{k}^{\mathrm{R}}),
\end{align}
where
\begin{align}
&\nabla_{\mathbf{c}_{k}^{\mathrm{R}}}\mathsf{h}_{l}(\mathbf{c}_{k}^{\mathrm{R}}) = \bigg[\frac{\partial\mathsf{h}_{l}(\mathbf{c}_{k}^{\mathrm{R}})}{\partial[\mathbf{c}_{k}^{\mathrm{R}}]_{1}}, \frac{\partial\mathsf{h}_{l}(\mathbf{c}_{k}^{\mathrm{R}})}{\partial[\mathbf{c}_{k}^{\mathrm{R}}]_{2}}\bigg]^{\mathsf{T}}, \\
&\frac{\partial\mathsf{h}_{l}(\mathbf{c}_{k}^{\mathrm{R}})}{\partial[\mathbf{c}_{k}^{\mathrm{R}}]_{1}} = -2\beta_{l}^{\mathrm{R}}|C_{k,l}^{\mathrm{R}}|F_{l}(\mathsf{sin}\vartheta_{\mathrm{R}}^{\mathrm{t}}\mathsf{sin}\varphi_{\mathrm{R}}^{\mathrm{t}} - \mathsf{sin}\vartheta_{\mathrm{R}}^{\mathrm{r}}\mathsf{sin}\varphi_{\mathrm{R}}^{\mathrm{r}}) \notag\\
&\qquad \times \sum_{j=1}^{J}\mathsf{sin}(F_{l}(\mathbf{c}_{k}^{\mathrm{R}} + \mathbf{t}_{k,j}^{\mathrm{R}})^{\mathsf{T}}(\bm{\rho}_{\mathrm{R}}^{\mathrm{t}} - \bm{\rho}_{\mathrm{R}}^{\mathrm{r}}) - \angle(C_{k,l}^{\mathrm{R}})), \\
&\frac{\partial\mathsf{h}_{l}(\mathbf{c}_{k}^{\mathrm{R}})}{\partial[\mathbf{c}_{k}^{\mathrm{R}}]_{2}} = -2\beta_{l}^{\mathrm{R}}|C_{k,l}^{\mathrm{R}}|F_{l}(\mathsf{cos}\varphi_{\mathrm{R}}^{\mathrm{t}} - \mathsf{cos}\varphi_{\mathrm{R}}^{\mathrm{r}}) \notag\\
&\qquad \times \sum_{j=1}^{J}\mathsf{sin}(F_{l}(\mathbf{c}_{k}^{\mathrm{R}} + \mathbf{t}_{k,j}^{\mathrm{R}})^{\mathsf{T}}(\bm{\rho}_{\mathrm{R}}^{\mathrm{t}} - \bm{\rho}_{\mathrm{R}}^{\mathrm{r}}) - \angle(C_{k,l}^{\mathrm{R}})), \\
&\mathbf{M}_{k,l}^{\mathrm{R}} \!\!=\!\! -2\beta_{l}^{\mathrm{R}}|C_{k,l}^{\mathrm{R}}|F_{l}^{2}J(\mathsf{Diag}([(\mathsf{sin}\vartheta_{\mathrm{R}}^{\mathrm{t}}\mathsf{sin}\varphi_{\mathrm{R}}^{\mathrm{t}} \!-\! \mathsf{sin}\vartheta_{\mathrm{R}}^{\mathrm{r}}\mathsf{sin}\varphi_{\mathrm{R}}^{\mathrm{r}})^{2}, \notag\\
&\qquad(\mathsf{cos}\varphi_{\mathrm{R}}^{\mathrm{t}} - \mathsf{cos}\varphi_{\mathrm{R}}^{\mathrm{r}})^{2}]^{\mathsf{T}}) + |(\mathsf{sin}\vartheta_{\mathrm{R}}^{\mathrm{t}}\mathsf{sin}\varphi_{\mathrm{R}}^{\mathrm{t}} - \mathsf{sin}\vartheta_{\mathrm{R}}^{\mathrm{r}}\mathsf{sin}\varphi_{\mathrm{R}}^{\mathrm{r}}) \notag\\
&\qquad \times (\mathsf{cos}\varphi_{\mathrm{R}}^{\mathrm{t}} - \mathsf{cos}\varphi_{\mathrm{R}}^{\mathrm{r}})|\mathbf{I}_{2}),
\end{align}
with $\bm{\Theta}_{k}(\mathbf{c}_{k}^{\mathrm{R}}) \triangleq \mathsf{Diag}(\mathbf{a}_{k}(\mathbf{c}_{k}^{\mathrm{R}}, f_{\mathrm{c}}, \vartheta_{\mathrm{R}}^{\mathrm{t}}, \varphi_{\mathrm{R}}^{\mathrm{t}}) \odot \mathbf{a}_{k}^{*}(\mathbf{c}_{k}^{\mathrm{R}}, f_{\mathrm{c}}, \vartheta_{\mathrm{R}}^{\mathrm{r}}, \varphi_{\mathrm{R}}^{\mathrm{r}}))$, $\beta_{l}^{\mathrm{R}} \triangleq \alpha_{\mathrm{G},l}^{2}\alpha_{\mathrm{h},l}^{2}(\mathsf{g}_{l}^{\mathrm{B}}(\tilde{\mathbf{p}}_{\mathrm{B}}))^{2}$, $C_{k,l}^{\mathrm{R}} \triangleq \sum_{n = 1, n \neq k}^{K}\mathbf{a}_{n}^{\mathsf{H}}(\mathbf{c}_{n}^{\mathrm{R}}, f_{l}, \vartheta_{\mathrm{R}}^{\mathrm{t}}, \varphi_{\mathrm{R}}^{\mathrm{t}})\bm{\Theta}_{n}(\mathbf{c}_{n}^{\mathrm{R}})\mathbf{a}_{n}(\mathbf{c}_{n}^{\mathrm{R}}, f_{l}, \vartheta_{\mathrm{R}}^{\mathrm{r}}, \varphi_{\mathrm{R}}^{\mathrm{r}})$ and $\bm{\rho}_{\mathrm{R}}^{\mathrm{t}} \triangleq [\mathsf{sin}\vartheta_{\mathrm{R}}^{\mathrm{t}}\mathsf{sin}\varphi_{\mathrm{R}}^{\mathrm{t}}, \mathsf{cos}\varphi_{\mathrm{R}}^{\mathrm{t}}]^{\mathsf{T}}$. Detailed derivations can be found in [15].

Next, we convexify (23c), which can be achieved by utilizing the following first-order Taylor expansion
\begin{align}
\| \mathbf{c}_{k}^{\mathrm{R}} - \mathbf{c}_{i}^{\mathrm{R}} \|_{2} \geq \frac{(\hat{\mathbf{c}}_{k}^{\mathrm{R}} - \mathbf{c}_{i}^{\mathrm{R}})^{\mathsf{T}}}{\| \hat{\mathbf{c}}_{k}^{\mathrm{R}} - \mathbf{c}_{i}^{\mathrm{R}} \|_{2}}(\mathbf{c}_{k}^{\mathrm{R}} - \mathbf{c}_{i}^{\mathrm{R}}).
\end{align}

By substituting (24) and (29) into the left hand sides of (23a) and (23c), respectively, we obtain the following convex optimization problem
\begin{align}
&(\mathcal{P}6_{k}): \max_{\mathbf{c}_{k}^{\mathrm{R}}, \kappa} \; \kappa \\
\mathrm{s.t.} \; &\mathsf{h}_{l}(\mathbf{c}_{k}^{\mathrm{R}}|\hat{\mathbf{c}}_{k}^{\mathrm{R}}) \geq \kappa, \; l \in \mathcal{L}, \tag{30a}\\
&\mathbf{c}_{k}^{\mathrm{R}} + \mathbf{t}_{k,j}^{\mathrm{R}} \in \mathcal{C}_{\mathrm{IRS}}, \; j \in \mathcal{J}, \tag{30b}\\
&\frac{(\hat{\mathbf{c}}_{k}^{\mathrm{R}} - \mathbf{c}_{i}^{\mathrm{R}})^{\mathsf{T}}}{\| \hat{\mathbf{c}}_{k}^{\mathrm{R}} - \mathbf{c}_{i}^{\mathrm{R}} \|_{2}}(\mathbf{c}_{k}^{\mathrm{R}} - \mathbf{c}_{i}^{\mathrm{R}}) \geq D_{\mathrm{IRS}}, \; i \in \mathcal{K}, \; k \neq i, \tag{30c}
\end{align}
which can be numerically solved by CVX.

The proposed algorithm for tackling $(\mathcal{P}1)$ is summarized in Algorithm 1.

\begin{algorithm}[!t]
\footnotesize
\caption{Solution to $(\mathcal{P}1)$}
\begin{algorithmic}[1]
\STATE Initialize feasible $\tilde{\mathbf{p}}_{\mathrm{B}}^{(0)}$, $\tilde{\mathbf{c}}_{\mathrm{R}}^{(0)}$ and $n = 0$;
\REPEAT
\FOR{$m = 1$ to $M$}
\STATE update $(\mathbf{p}_{m}^{\mathrm{B}})^{(n + 1)}$ by solving $(\mathcal{P}4_{m})$;
\ENDFOR
\FOR{$k = 1$ to $K$}
\STATE update $(\mathbf{c}_{k}^{\mathrm{R}})^{(n + 1)}$ by solving $(\mathcal{P}6_{k})$;
\ENDFOR
\STATE $n := n + 1$;
\UNTIL{convergence}
\end{algorithmic}
\end{algorithm}

\section{Simulation Results}

\begin{figure}[!t]
\centering
\includegraphics[scale=0.45]{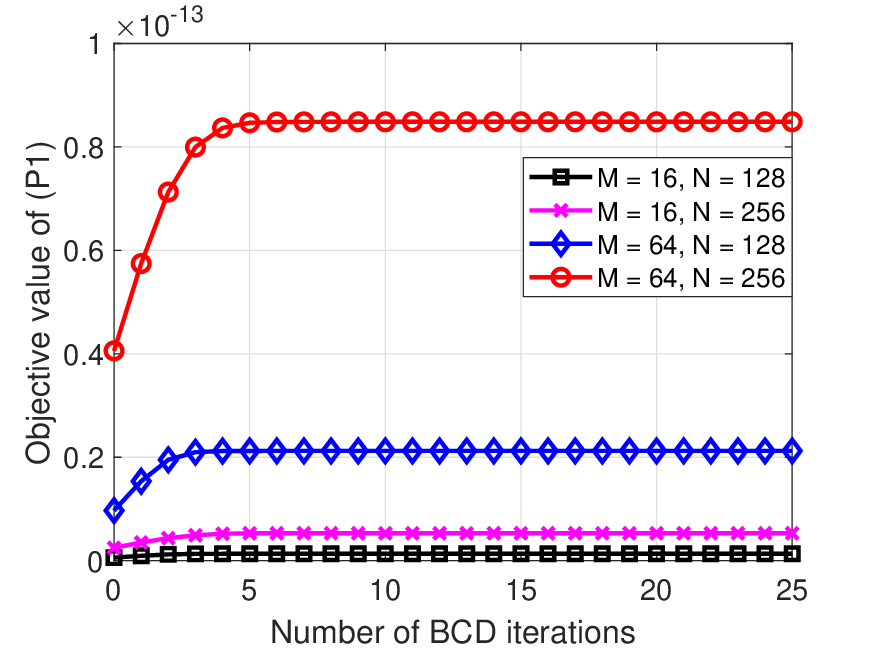}
\caption*{Fig. 2. Convergence behaviour of Alg. 1.}
\end{figure}

\begin{figure}[!t]
\centering
\includegraphics[scale=0.45]{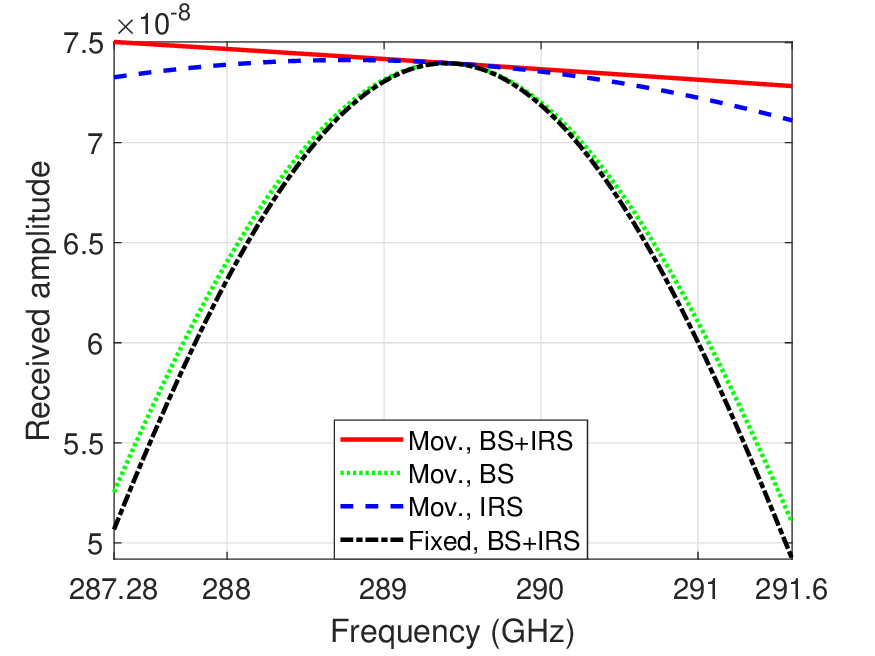}
\caption*{Fig. 3. Received amplitude across considered THz frequency band.}
\end{figure}

In this section, we provide numerical results to demonstrate the effectiveness of our proposed algorithm and the benefit of deploying antenna array and IRS with movable components. Unless otherwise specified, $M = 16$, $N = 256$, $K = N$ (i.e., $J = J_{1}J_{2} = 1 \times 1 = 1$), $L = 128$, $f_{0} = 287.28\mathrm{GHz}$, $f_{L} = 291.60\mathrm{GHz}$, $f_{\mathrm{c}} = \frac{f_{0} + f_{L}}{2} = 289.44\mathrm{GHz}$ [16], $f_{l} = f_{0} + \frac{l}{L}(f_{L} - f_{0}), \; l \in \mathcal{L}$, $A_{\mathrm{BS},1} = A_{\mathrm{BS},2} = \frac{25c}{f_{c}}$, $A_{\mathrm{IRS},1} = A_{\mathrm{IRS},2} = \frac{50c}{f_{c}}$, $D_{\mathrm{BS}} = \frac{c}{2f_{c}}$, $D_{\mathrm{IRS}} = (1 + \sqrt{(J_{1} - 1)^{2} + (J_{2} - 1)^{2}})\frac{c}{2f_{c}}$, $\kappa_{\mathrm{abs}}(f_{l}) = 5.157 \times 10^{-4} \mathrm{dB/m}$.

Fig. 2 illustrates the convergence behaviour of our proposed Alg. 1. As suggested by Fig. 2, by utilizing Alg. 1 which starts from feasible initial points, the objective value of $(\mathcal{P}1)$ guarantees monotonically increasing characteristics and converges in several (lower than $10$) iterations.

Fig. 3 plots the amplitude of user's received signal (defined in (9)) across different subcarriers operating in THz frequency band. As shown in Fig. 3, the double beam squint effect can be completely eliminated via appropriately configuring the positions of MAs and subarrays. Besides, with the double beam squint effect mitigated, the received amplitude is not equal across considered frequency band. The reason lies in that, as shown by (9), the received amplitude is related to path loss, which becomes smaller when subcarrier's frequency increases.

\section{Conclusion}

In this paper, a wideband THz MISO system aided by an MA array and an IRS with movable subarrays is considered. To overcome severe double beam squint effect, a minimal received power maximization problem is formulated to configure MAs' and IRS subarrays' positions. Based on MM framework, we successfully develop an algorithm to resolve this challenging optimization. Simulation results verify that double beam squint effect can be significantly suppressed by adjusting the positions of movable components within the BS and IRS.

\end{document}